\documentclass[
fleqn,usenatbib]{mnras}

\usepackage{newtxtext,newtxmath}

\usepackage[T1]{fontenc}
\usepackage{ae,aecompl}

\usepackage{graphicx}	
\usepackage{amsmath}	
\usepackage{amssymb}	
\usepackage{multirow}   

\newcommand{\omk}{\Omega_K}
\newcommand{\omm}{\Omega_m}
\newcommand{\lcdm}{$\Lambda$CDM }
\newcommand{\lcdmk}{$\Lambda\text{CDM}+\Omega_K$}

\allowdisplaybreaks[4] 

\title[Cosmography with Spatial Curvature]{General Cosmography Model with Spatial Curvature}

\author[E. K. Li et al.]{
En-Kun Li,\thanks{E-mail: ekli\_091@mail.dlut.edu.cn}
Minghui Du,\thanks{E-mail: angelbeats@mail.dlut.edu.cn}
Lixin Xu\thanks{Corresponding author (L. Xu), E-mail: lxxu@dlut.edu.cn}
\\
Institute of Theoretical Physics, School of Physics, Dalian University of Technology, Dalian, 116024, P. R. China
}

\date{Accepted XXX. Received YYY; in original form ZZZ}

\pubyear{2015}

\begin{document}
\label{firstpage}
\pagerange{\pageref{firstpage}--\pageref{lastpage}}
\maketitle

\begin{abstract}
    The cosmographic approach is adopted to determine the spatial curvature (i.e., $\Omega_K$) combining the latest released cosmic chronometers data (CC), the Pantheon sample of type Ia supernovae observations, and the baryon acoustic oscillation measurements.
    We use the expanded transverse comoving distance $D_M(z)$ as a basic function for deriving $H(z)$ and the other cosmic distances.
    In this scenario, $\Omega_K$ can be constrained only by CC data.
    To overcome the convergence issues at high-redshift domains, two methods are applied: the Pad\'{e} approximants and the Taylor series in terms of the new redshift $y=z/(1+z)$.
    Adopting the Bayesian evidence, we find that there is positive evidence for the Pad\'{e} approximant up to order ($2,2$) and weak evidence for the Taylor series up to 3-rd order against $\Lambda\text{CDM}+\Omega_K$ model.
    The constraint results show that a closed universe is preferred by the present observations under all the approximants used in this study.
    And the tension level of the Hubble constant $H_0$ is less than $2\sigma$ significance between different approximants and the local distance ladder determination.
    For each assumed approximant, $H_0$ is anti-correlated with $\Omega_K$ and the sound horizon at the end of the radiation drag epoch, which indicates that the $H_0$ tension problem can be slightly relaxed by introducing $\Omega_K$ or any new physics which can reduce the sound horizon in the early universe.
\end{abstract}




\section{Introduction}
\label{sec:intr}

Nowadays, a huge number of independent observations provide strong evidence to support a late-time accelerated expanding universe \citep{Ade:2015xua, Aghanim:2018eyx}.
This observed phenomenon is one of the major puzzles in modern cosmology.
In general, there are two kinds of interpretations for this cosmic phase: i) postulating an exotic form of energy with negative pressure usually called dark energy or ii) modifying the laws of gravity.
Numerous models have been proposed based on these two branches, but it is difficult to determine which one is correct due to the degeneracies in the parameter space. 
Despite this, the Lambda Cold Dark Matter (\lcdm) model with six parameters could excellently fit almost all observational data, and has been set as the standard model of cosmology \citep{Ade:2015rim}.

By assuming flat \lcdm model, the final full-mission \textit{Planck} measurements of the cosmic microwave background (CMB) anisotropies has found a value of the Hubble constant $H_0=(67.27\pm0.60)$~km/s/Mpc \citep{Aghanim:2018eyx}.
This is compatible with many earlier and recent estimates of $H_0$ \citep{Gott:2000mv, Chen:2003gt, Chen:2011ab, Chen:2016uno, Wang:2017yfu, Lin:2017bhs, Abbott:2017smn, Haridasu:2018gqm, Zhang:2018ida, Zhang:2018air, Dominguez:2019jqc}.
In contrast, multiple local expansion rate measurements find slightly higher $H_0$ values and slightly larger error bars \citep{Rigault:2014kaa, Zhang:2017aqn, Dhawan:2017ywl, Fernandez-Arenas:2017isq}.
And the latest value from the 
Supernovae and HO for the Dark Energy Equation of State (SH0ES) project 
together with GAIA DR2 parallaxes is $H_0=(73.52\pm 1.62)$~km/s/Mpc (hereafter R18), which has a more than $3\sigma$ tension with the \textit{Planck} CMB data \citep{Riess:2018byc}.
This tension is one of the most intriguing problems in modern cosmology.
There have been many attempts to solve the problem, such as introducing new physics beyond the standard \lcdm cosmological model \citep{Wyman:2013lza, Zhao:2017urm,DiValentino:2017oaw, DiValentino:2017rcr, Sola:2017znb, Yang:2018uae}, reanalyzing of the SH0ES data \citep{Efstathiou:2013via, Cardona:2016ems, Zhang:2017aqn, Follin:2017ljs, Feeney:2017sgx}, etc.

In the present paper, we are interested in figuring out whether the present distance observations prefer a non-zero spatial curvature or not, and how the spatial curvature might be tied in with the $H_0$ tension problem.
Current constraint on the spatial curvature parameter $\omk$ by the combination of the only \textit{Planck} CMB data within the \lcdmk model is $\omk = -0.044^{+0.018}_{-0.015}$, which favors a closed universe model at more than $2\sigma$ confidence level \citep{Aghanim:2018eyx}.
However, adding lensing and baryon acoustic oscillation measurement (BAO) data pulls parameters back into consistency with a spatially flat universe ($\omk = 0.0007 \pm 0.0019$) \citep{Aghanim:2018eyx}.
Apart from the Hubble scale, non-zero spatial curvature sets an additional length scale, and thereby assuming a power law spectrum for energy density inhomogeneities in non-flat models is incorrect \citep{Ratra:1994vw, Ratra:2017ezv}.
However, the non-flat slow-roll inflation models \citep{Gott:1982zf, Hawking:1983hj, Ratra:1984yq} provide the physically consistent mechanism for generating energy density inhomogeneities in the non-flat case.
The power spectra in these models have been computed in Ref. \citep{Ratra:1994vw, Ratra:2017ezv}.
If one use these untilted non-flat inflation power spectra in the analysis of the CMB data, a mildly closed universe is favored \citep{Ooba:2017ukj, Ooba:2017npx, Ooba:2017lng, Park:2017xbl, Park:2018bwy, Park:2018fxx, Park:2019emi}.
Additionally, there are also some evidences for non-flat geometries \citep{Farooq:2013dra, Farooq:2016zwm, Rana:2016gha, Ryan:2018aif, Park:2018tgj, Abbott:2018xao, Zheng:2019trp, Cheng-Zong:2019iau, Ryan:2019uor, Handley:2019anl, Khadka:2019njj}, which are given under various combinations of cosmic observations, such as BAO, type Ia supernovae observations (SNe Ia), observational Hubble data $H(z)$, redshift space distortion measurements, weak lensing, etc.

We should note that modern cosmology is based on the Friedmann equations.
But the Hubble relation between distance and redshift is a purely cosmographic relation that depends only on the symmetry of the Friedmann-Lema\^{i}tre-Robertson-Walker (FLRW) spacetime.
And it does not intrinsically require any dynamical assumptions \citep{Cattoen:2008th}. 
This suggests that it should be possible to characterize the late-time cosmic expansion with purely kinetic parameters based on the Hubble relation. 

Instead of particular dynamical cosmological models, one can use a model-independent kinematic approach called cosmography \citep{Chiba:1998tc, Visser:2003vq, Visser:2004bf, Capozziello:2013wha, Dunsby:2015ers} to describe the evolution of the Hubble parameter and cosmic distances.
The only assumption of the purely kinematic approach is the cosmological principle, i.e., the FLRW metric.
And the parameters in a cosmography model can be used to determine the kinematical status of our universe.
For example, 
the Hubble constant $H_0$ describes the current expansion rate of our universe, and the current deceleration parameter $q_0$ describes whether our universe is experiencing accelerated expansion.
Until now, the cosmography method has been widely used in studying the modified gravity theories \citep{Aviles:2012ir, Aviles:2013nga}, the features of dark energy \citep{Luongo:2013rba, Luongo:2015zgq}, and the different cosmographic parameters \citep{Xu:2010hq, Luongo:2011zz, Aviles:2012ay, Aviles:2016wel}, etc.

Owing to the absence of additional physical assumptions, purely geometrical and model-independent methods may be better at measuring spatial curvature.
Several model-independent methods have been proposed to determine the spatial curvature parameter $\omk$, such as adopting the sum rule of distances along null geodesics of the FLRW metric \citep{Bernstein:2005en, Rasanen:2014mca, Liao:2017yeq, Xia:2016dgk, Li:2018hyr, Qi:2018aio}, combining the Hubble parameter $H(z)$, the transverse comoving distance $D_M(z)$ \citep{Hogg:1999ad} and its derivation with redshift $D_M'(z)$ \citep{Clarkson:2007pz, Clarkson:2007bc, Shafieloo:2009hi, Sapone:2014nna, Li:2014yza, Yahya:2013xma, Cai:2015pia, LHuillier:2016mtc, Rana:2016gha}, 
and deriving the $\omk$ by comparing the distance derived from $H(z)$ and $D_M(z)$ \citep{Yu:2016gmd}, which is further applied to new data-set to constrain $\omk$ \citep{Li:2016wjm, Wang:2017lri, Wei:2016xti, Yu:2017iju}, etc.
These investigations suggest that the non-zero $\omk$ cannot be ruled out by the current observations.

The layout of our paper is as follows: 
In section~\ref{sec:cosmography}, we introduce the general cosmography model with spatial curvature and raise our new cosmography model. 
The data-set used in this analysis and the main methodology are described in section~\ref{sec:obs_meth}.
Our results and analysis are presented in section~\ref{sec:results}.
We summarize our conclusions in section~\ref{sec:conclusions}.

\section{Cosmographic Model with Spatial Curvature}
\label{sec:cosmography}

Recently, the cosmographic approach , which preserves the minimum priors of isotropy and homogeneity while ignoring other assumptions, has gained increasing interest in capturing as much information as possible directly from cosmic observations.
Actually, the only assumption retained in this approach is the FLRW metric
\begin{equation}
    ds^2=-c dt^2+a^2(t)\left[\frac{dr^2}{1-K r^2}+r^2(d{\theta}^2+\sin^2{\theta}d{\phi}^2)\right],
    \label{eq:FLRW}
\end{equation}
where $c$ is the speed of light, the parameter $K=1,0,-1$ denote the spatial curvature for closed, flat and open geometries, respectively.
The cosmographic approach starts by defining the cosmographic functions \citep{Chiba:1998tc, Visser:2003vq, Dabrowski:2004hx, Dabrowski:2005fg},
\begin{eqnarray}
    H = \frac{\dot{a}}{a}, ~~ q= -\frac{\ddot{a}}{aH^2}, ~~ j= \frac{a^{(3)}}{aH^3}, ~~ s = \frac{a^{(4)}}{aH^4}, ~~ l = \frac{a^{(5)}}{aH^5}, \cdots
    \label{eq:hqjsl}
\end{eqnarray}
where the dots represent cosmic time derivatives and $a^{(n)}$ stands for the $n$-th time derivative of $a$.
The present values of these functions are Hubble constant, deceleration, jerk, snap and lerk parameters, respectively.
    
\subsection{The general cosmography model}
\label{sec:gen_cosgr}

Now considering the path of a photon ($ds=0$), one has
\begin{equation}
  c dt = a(t) \frac{dr}{\sqrt{1-Kr^2}}.
\end{equation}
Integrating the equation, one can obtain the comoving distance
\begin{equation}
  d_c \equiv a_0 \int_0^{r_e} \frac{dr}{\sqrt{1-K r^2}} = -a_0 \int_{t_e}^{t_0} \frac{c dt}{a(t)},
  \label{eq:com_dc}
\end{equation}
where $t_e$ is the time when the photon was emitted at $r=r_e$, and $t_0$ is the time when it was observed at $r=0$.
Substituting $a/a_0 = 1/(1+z)$, then the comoving distance is given by
\begin{equation}
  d_c = \frac{c}{H_0} \int_0^z \frac{dz'}{E(z')},
\end{equation}
where $E(z)\equiv H(z)/H_0$.
From Eq.~\eqref{eq:com_dc}, one can also derive the transverse comoving distance 
\begin{equation}
  D_M(z) =
  \begin{cases}
    \frac{c}{H_0 \sqrt{\omk}} \sinh\left( \frac{H_0 \sqrt{\omk}}{c} d_c \right), & K = -1, \\
    d_c, & K = 0, \\
    \frac{c}{H_0 \sqrt{\omk}} \sin\left( \frac{H_0 \sqrt{\omk}}{c} d_c \right), & K = +1,
  \end{cases}
  \label{eq:dm_def}
\end{equation}
where $\omk = -Kc^2/(a_0 H_0)^2$ is the spatial curvature parameter.
Then, using the definitions of cosmographic parameters in Eq.~\eqref{eq:hqjsl}, $D_M(z)$ can be expanded up to the 5-th order 
\begin{equation}
    D_M(z) = \frac{c}{H_0} \sum_{i=1} d_i z^i,
    \label{eq:dm_in_z}
\end{equation}
where
\begin{align}
    d_1 &= 1, \\
    d_2 &= -\frac{1}{2} \left( 1+ q_0 \right), \\
    d_3 &= \frac{1}{6} \left( 2 +4q_0 +3q_0^2 -j_0 + \omk \right), \\
    d_4 &= -\frac{1}{24} \left( 6 +18 q_0-27 q_0^2-15 q_0^3 -j_0 (9+10 q_0) -s_0 \right. \nonumber \\
    &~~ \left. +6 (1+q_0) \omk \right), \\
    d_5 &= \frac{1}{120} \left( 24+10 j_0^2-l_0+96 q_0+216 q_0^2+240 q_0^3+105 q_0^4 \right. \nonumber \\
    & ~~ \left. -j_0 (72+160 q_0+105 q_0^2) -16 s_0-15 q_0 s_0 \right. \nonumber \\
    & ~~ \left. +5 (7-2 j_0+14 q_0+9 q_0^2) \omk +\omk^2 \right).
\end{align}

However, cosmography in this form encounters convergence problems at high redshifts ($z> 1$). 
To solve this trouble,  $y$-redshift is hence introduced \citep{Cattoen:2007sk}
\begin{equation}
    y = \frac{z}{1+z}.
    \label{eq:z2y}
\end{equation}
Taylor series in the $y$-redshift are likely to be well behaved at high redshift region and then many cosmic observations such as SNe Ia with higher redshifts can be used to fit the cosmography model.

Nevertheless, the expansion in $y$ still has some highly undesirable properties. 
Firstly, since $y$ tends to unity when $z$ tends to infinity, $H(z)$ or $d(z)$  tends to a constant at high redshift, which is a restriction that is both unnecessary and conflicting with our present understanding of the universe. 
Secondly, the Taylor series dose not converge when $y<-1$ (namely $z<-1/2$), and it drastically diverges when $z\rightarrow -1$ \citep{Gruber:2013wua}.

To solve or alleviate this problem of original cosmography, some generalizations of cosmography are introduced, such as Taylor series in terms of different functions of redshift $z$ \citep{Cattoen:2007sk, Aviles:2012ay}, applying the Pad\'{e} approximant \citep{Gruber:2013wua, Aviles:2014rma}, and using the ratios of Chebyshev polynomials method \citep{Capozziello:2017nbu}, etc.
In the present paper, we will use the Pad\'{e} approximant to solve this problem.
For a function $f(z)$, the Pad\'{e} approximant of order ($m,n$) is given by the rational function \citep{pade1892representation, Gruber:2013wua, Aviles:2014rma}
\begin{equation}
  P_{mn}(z) = \frac{\alpha_0 +\alpha_1 z +\alpha_2 z^2 + \cdots +\alpha_m z^m}{1 +\beta_1 z + \beta_2 z^2 +\cdots + \beta_n x^n},
  \label{eq:Pade}
\end{equation}
where $m$ and $n$ are non-negative integers, and $\alpha_i$, $\beta_i$ are constant and should agree with $f(z)$ and its derivatives at $z = 0$ to the highest possible order, i.e. such that $P_{mn}(0) = f(0)$, $P_{mn}^\prime(0) = f^\prime(0)$, $\cdots$, $P_{mn}^{(m+n)}(0) = f^{(m+n)}(0)$.
Obviously, it reduces to the Taylor series when all $\beta_i=0$ or $n=0$.
The Pad\'{e} approximant often gives a better approximation of the function than that of truncating its Taylor series, and it may still work at the points where the Taylor series does not converge.
Thus,  the Pad\'{e} approximant of $D_M(z)$ can be constructed from Eq.~\eqref{eq:dm_in_z}.

Specially, from Eq.~\eqref{eq:dm_def} one can find that $D_M(0)=0$ and $D_M^\prime(0) = c/H_0$, then the Pad\'{e} approximant of $H_0 D_M(z)/c$ can be written as
\begin{equation}
    P^d_{mn}(z) = \frac{z +\alpha_2 z^2 + \cdots +\alpha_m z^m}{1 +\beta_1 z + \beta_2 z^2 +\cdots + \beta_n x^n}.
    \label{eq:Pade_d}
\end{equation}
Moreover, in order to include $\omk$ in $D_M(z)$, there should be at least one of the Pad\'{e} approximant order ($m,n$)  bigger than 2.
Then, we have the following different orders of approximation: 1). the polynomial of thrid degree, i.e. the Pad\'{e} approximants of degree ($2,1$) and ($1,2$); 2). the polynomial of fourth degree, i.e. Pad\'{e} approximants of degree ($3,1$), ($2,2$) and ($1,3$); 3). the polynomial of fifth degree, i.e. Pad\'{e} approximants of degree ($4,1$), ($3,2$), ($2,3$) and ($1,4$).
Some explicit expressions are reported in Appendix~\ref{sec:app_A}. 

Here, if we directly expand $H(z)$ in terms of $z$ or $y$, one can find that there is no curvature parameter in the Hubble parameter function.
While within this scheme, $\omk$ acts as a nuisance parameter for the cosmic chronometers data (please refer to section~\ref{sec:ohd} for details), which will surely depress the usefulness of the data-set and result in poor constraining performance.
So it is a great necessity to find a suitable expression for $H(z)$ with $\omk$ included.
From Eq.~(\ref{eq:dm_def}), it can be obtained that the Hubble function in terms of $\omk$ and $z$ is
\begin{align}
    H(z, \omk) &= \frac{c}{\partial D_M(z)/\partial z}\sqrt{1+ \frac{H_0^2 \Omega _K}{c^2} D_M(z)^2 }.
    \label{eq:HofDM}
\end{align}
Meanwhile, the luminosity distance and angular diameter distance are $D_L(z) = (1+z) D_M(z)$ and $D_A(z) = D_M(z)/(1+z)$, respectively.
Then, using $D_M(z)$ one can reconstruct the luminosity distance, the angular diameter distance, and the Hubble parameter function with $\omk$.

\subsection{\lcdmk model with cosmographic parameters}
\label{sec:rebuilt_lcdm}

In order to give a qualitative representation of the Pad\'{e} approximation of $D_M(z)$, we compare the numerical behavior of different approximations with the \lcdmk model over a large range of $z$.
The Hubble parameter in \lcdmk model reads
\begin{equation}
    H = H_0 \sqrt{\omm(1+z)^3 +\omk (1+z)^2 +(1-\omm-\omk)},
    \label{eq:lcdm_omk}
\end{equation}
and the corresponding cosmographic parameters are
\begin{align}
    q_0 &= \frac{3}{2}\omm +\omk -1, \label{eq:q0}\\
    j_0 &= 1-\omk, \label{eq:j0}\\
    s_0 &= 1-\frac{9}{2}\omm +\omk^2 -\omk \left(2-\frac{3}{2}\omm \right), \label{eq:s0}\\
    l_0 &= 1+3\omm +\frac{27}{2}\omm^2 +\omk^2 -\omk (2-9\omm), \label{eq:l0}
\end{align}
where $\omm$ and ($1-\omm-\omk$) represent the energy density of dark matter and dark energy, respectively.
Here $\omm$ and $\omk$ can be rewritten in terms of $q_0$ and $j_0$ as
\begin{align}
    \omm &= \frac{2}{3} \left(q_0 +j_0\right), \\
    \omk &= 1-j_0,
\end{align}
and then
\begin{align}
    s_0 &= -2(q_0+j_0)-q_0 j_0, \\
    l_0 &= j_0^2 +2(q_0 +j_0)(4+3q_0).
\end{align}

\subsection{Performance of the cosmography models}
\label{sec:display}

In general, higher order of expansions could provide more accurate approximations.
However, in this way, more model parameters need to be introduced.
This raises another problem, which is how many series terms we need to include to obtain a good approximation of the model functions.
This question had been discussed in Refs.~\citep{Cattoen:2008th, Vitagliano:2009et, Xia:2011iv} by the $F$-test method and in Ref.~\citep{Zhang:2016urt} by the $Risk$ method.
In the present paper, the Bayesian evidence method will be adopted to study this question in section~\ref{sec:bayesian}.

\begin{figure*}
    \centering
    \includegraphics[scale=0.55]{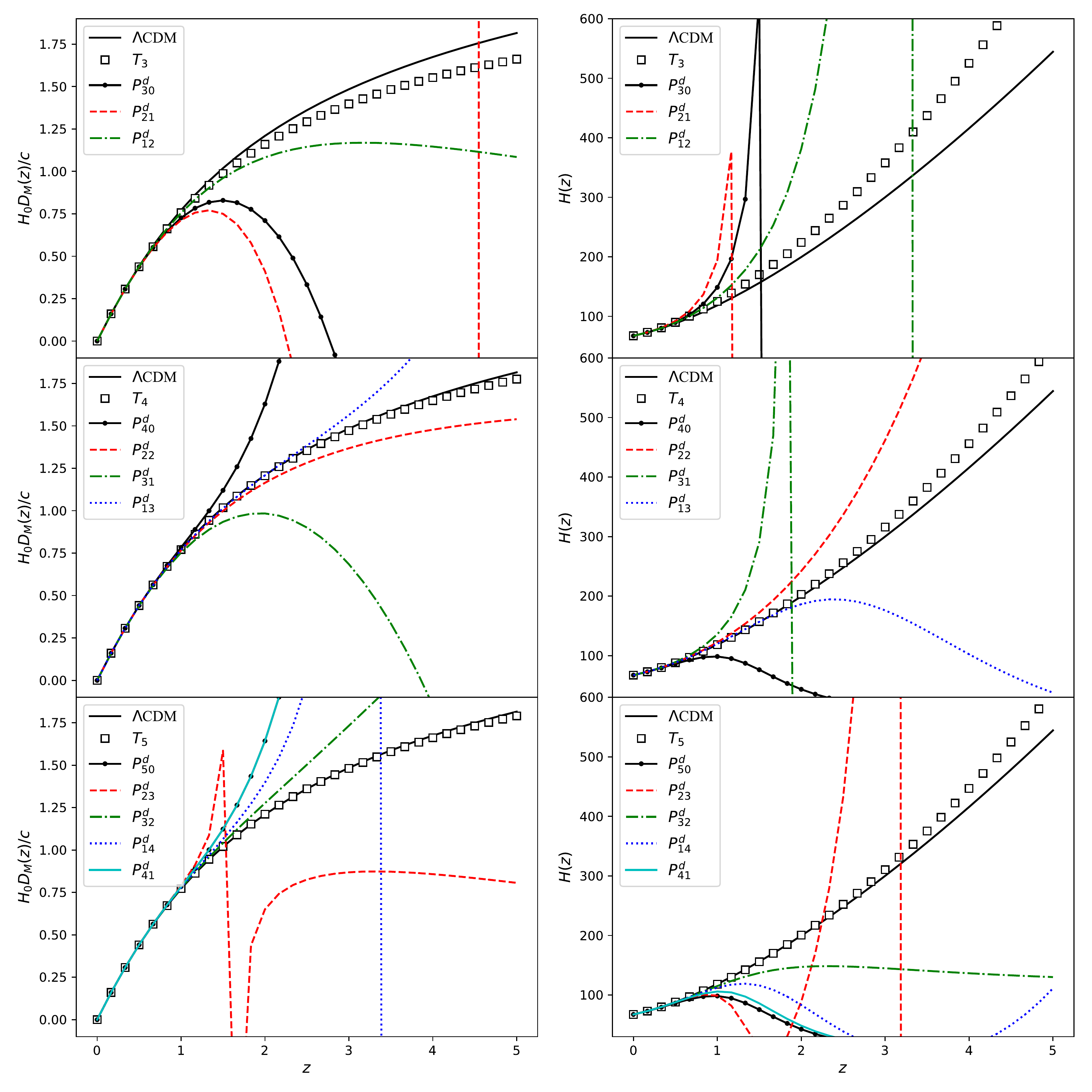}
    \caption{Analytical curves for the dimensionless distance $d(z)$ and Hubble parameter $H(z)$ of the \lcdm model compare to the Taylor and Pad\'{e} approximations. Here we use the parameters $H_0 = 67.27$, $q_0 = -0.55$, $j_0=1$, $s_0 = -0.35$, and $l_0 = 3.115$, which are calculated using Eqs.~\eqref{eq:q0} - \eqref{eq:l0} with $\omm =0.3$ and $\omk=0.0$.}
    \label{fig:pade_dh}
\end{figure*}

At the end of this section, we would like to present the qualitative features of the different Pad\'{e} and Taylor approximant models, and their deviation from the \lcdm model.
The plots of the dimensionless distance $d(z) = H_0 D_M(z)/c$ and Hubble parameter $H(z)$ for different orders of approximant are shown in Fig.~\ref{fig:pade_dh}.
Here, $P_{mn}^d$ represents the Pad\'{e} approximant of degree ($m,n$), and $T_i$ is the Taylor polynomial of $i$-th order expansion in terms of $y=z/(1+z)$.
From the graphics in Fig.~\ref{fig:pade_dh}, one can immediately notice that the Pad\'{e} approximants (3,0), (4,0) and (5,0) or the general Taylor series in terms of $z$ (black line marked by dot in the figures) are accurate at small $z$ and quickly diverge from \lcdm model outside the region $z<1.5$.
However, the Taylor series in terms of $y$ (black square in the figures), i.e., $T_3$, $T_4$ and $T_5$, give excellent approximations to $d(z)$ and $H(z)$ of the \lcdm model.
At the same time, $P_{12}^d$, $P_{22}^d$, $P_{13}^d$ and $P_{32}^d$ also give good approximations to the \lcdm model over the interval considered in the following section ($z<2.4$).

Given this fact, in the following sections we will give a quantitative analysis of the different Pad\'{e} approximations ($P_{12}^d$, $P_{22}^d$, $P_{13}^d$, and $P_{32}^d$ cases) and Taylor polynomials in terms of $y$ ($T_3$, $T_4$, and $T_5$) for $D_M(z)$, by comparing them with the present cosmological observational combinations.
In this way we can get the optimal values of the cosmographic parameters by using the Markov Chain Monte Carlo sampling method.

\section{Observations and Methodology}
\label{sec:obs_meth}

To understand the degeneracies and best-fit of the cosmographic parameters, we need to sample different parameters using the currently available observational data-set.
In this section, we present the relevant observational data and the fitting methodology used to constrain the cosmography model.
In what follows, we will give a brief review about them.

\subsection{Observational Hubble Data}
\label{sec:ohd}
    
The $H(z)$ measurements can be obtained in two ways.
One way is based on the clustering of galaxies or quasars, which was firstly proposed by \cite{Gaztanaga:2008xz} using the BAO peak positions as the standard ruler in the radial direction.
However, some $H(z)$ data points in this method are biased due to the sound horizon $r_s$ is estimated by assuming a fiducial cosmological model (see subsection~\ref{sec:BAO}).

Another method comes from calculating the differential ages of passively evolving galaxies at different redshifts, providing $H(z)$ measurements that are model-independent \citep{Jimenez:2001gg}.
In this method, a change rate $\Delta z/\Delta t$ can be obtained, then the Hubble parameter $H(z)$ could be written as
\begin{equation}
    H(z) \simeq - \frac{1}{1+z} \frac{\Delta z}{\Delta t}.
    \label{eq:Hubble}
\end{equation}
This method is often referred to as cosmic chronometers.
In this paper, we select 31 cosmic chronometers $H(z)$ points (CC) in our model-independent analysis, and the current measurements of $H(z)$ are summarized in Table~\ref{tab:Hubble}.

\begin{table}
    \centering
    \caption{The latest Hubble parameter measurements $H(z)$ (in units of km/s/Mpc) and their errors $\sigma_H$ at redshift $z$ from the differential age method.}
    \label{tab:Hubble}
    \begin{tabular}{lllll}
        \hline
        $z$ & $H(z)$ & $\sigma_H$ & Reference \\
        \hline
        0.07   &  69.0  &  19.6   & \multirow{4}*{\citep{Zhang:2012mp}} \\
        0.12   &  68.6  &  26.2   & \\
        0.2    &  72.9  &  29.6   & \\
        0.28   &  88.8  &  36.6   & \\
        \hline
        0.1    &  69    &  12     & \multirow{11}*{\citep{Stern:2009ep}} \\
        0.17   &  83    &  8      & \\
        0.27   &  77    &  14     & \\
        0.4    &  95    &  17     & \\
        0.48   &  97    &  60     & \\
        0.88   &  90    &  40     & \\
        0.9    &  117   &  23     & \\
        1.3    &  168   &  17     & \\
        1.43   &  177   &  18     & \\
        1.53   &  140   &  14     & \\
        1.75   &  202   &  40     & \\
        \hline
        0.1797 & 75     & 4       & \multirow{8}*{\citep{Moresco:2012jh}} \\
        0.1993 & 75     & 5       & \\
        0.3519 & 83     & 14      & \\
        0.5929 & 104    & 13      & \\
        0.6797 & 92     & 8       & \\
        0.7812 & 105    & 12      & \\
        0.8754 & 125    & 17      & \\
        1.037  & 154    & 20      & \\
        \hline
        0.3802 & 83     & 13.5    & \multirow{5}*{\citep{Moresco:2016mzx}} \\
        0.4004 & 77     & 10.2    & \\
        0.4247 & 87.1   & 11.2    & \\
        0.4497 & 92.8   & 12.9    & \\
        0.4783 & 80.9   & 9       & \\
        \hline
        1.363  & 160    & 33.6    & \multirow{2}*{\citep{Moresco:2015cya}} \\
        1.965  & 186.5  & 50.4    & \\
        \hline
        0.47   & 89     & 34      & \citep{Ratsimbazafy:2017vga} \\
        \hline
    \end{tabular}
\end{table}

The best-fit parameters are obtained by minimizing the quantity
\begin{equation}
    \chi_\text{CC}^2(\{\theta_i\}) \equiv \sum\limits_{i} \frac{\left[ H(z_i|\{\theta_i\}) - H^{obs}(z_i) \right]^2 }{\sigma_{H,i}^2},
    \label{eq:chiH}
\end{equation}
where $H^{obs}(z_i)$ are the 31 CC $H(z)$ at $z_i$ and $\sigma_{H,i}^2$ is the measurement error of the $H^{obs}(z_i)$.

\subsection{Supernovae Ia Data}
\label{sec:JLA}

SNe Ia are widely accepted as the standard candles to measure the cosmological luminosity distance.
From the observational point of view, the observed distance modulus of each SNe is given by
\begin{equation}
    \mu_{\rm SN} = m_{\rm B}^* -M' = m_{\rm B}^* - (M_{\rm B} - \alpha X_1 + \beta \mathcal{C}),
    \label{eq:muSN}
\end{equation}
where $m_{\rm B}^*$ is the observed peak magnitude in rest frame B-band, $X_1$ is the time stretching of the light-curve, $\mathcal{C}$ is the SNe color at maximum brightness, $M_{\rm B}$ is the absolute magnitude.
And $\alpha, \beta$ are two nuisance parameters, which should be fitted simultaneously with the cosmological parameters.
However, this method strongly depends on a specific cosmological model.
To avoid this, \cite{Kessler:2016uwi} proposed a new method called BEAMS with Bias Corrections (BBC) to calibrated the SNe, and the corrected apparent magnitude $m_{\rm B,corr}^* = m_{\rm B}^* + \alpha X_1 -\beta \mathcal{C} +\Delta_B$ for all the SNe is reported in Ref. \citep{Scolnic:2017caz}, where $\Delta_B$ is the correction term.
The new data-set called Pantheon sample contains 1048 SNe spanning the redshift range $0.01<z<2.3$ \citep{Scolnic:2017caz}.
This is the largest spectroscopically confirmed SNe Ia sample released to date.

The theoretical distance modulus $\mu(z)$ is defined as
\begin{equation}
    \mu_{th}(z) = 5\log_{10}\left[ \frac{D_L(z)}{\rm Mpc} \right] +25 = 5\log_{10}[d_L(z)] +\mu_0,
\end{equation}
where $d_L(z) = H_0 D_L(z)$ is the Hubble-free luminosity distance.
Thus, likelihood for the Pantheon data is
\begin{align}
    \chi_\text{Pantheon}^2(\{\theta_i\}, M') \equiv {\bf \Delta \hat{\mu}^T} \cdot \textbf{Cov}^{-1} \cdot {\bf \Delta \hat{\mu}} 
\end{align}
where $\Delta \hat{\mu}_i = m_{{\rm B},i}^* - 5\log_{10}[d_L(z_i)] +(M'-\mu_0)$, and $M'-\mu_0$ can be marginalized over analytucally \citep{DiPietro:2002cz,Nesseris:2004wj, Nesseris:2005ur}.

\subsection{Baryon Acoustic Oscillation Data}
\label{sec:BAO}

Another key tool to probe the expansion rate and the large-scale properties of the universe is the BAO data, which is the imprint in the large-scale structure of matter due to the oscillations in the primordial plasma.
The BAO data used in this paper includes the measurements from the 6dF Galaxy Survey (6dFGS) \citep{Beutler:2011hx}, the Sloan Digital Sky Survey Data Release 7 (SDSS DR7) main galaxy sample (MGS) \citep{Ross:2014qpa}, the BOSS DR12 consensus BAO measurements \cite{Alam:2016hwk}, the extended BOSS (eBOSS) DR14 measurements from quasars clustering \citep{Ata:2017dya,Hou:2018yny}, the eBOSS DR11 cross-correlations of the
Ly$\alpha$ absorption with the distribution of quasars \cite{Font-Ribera:2013wce}, and SDSS DR12 Ly$\alpha$ forest \citep{Bautista:2017zgn}.
The 12 data points are summarized in Table.~\ref{tab:BAO}.

We note that the likelihood for the MGS data used in this work can not be well approximated by a Gaussian.
Thus, we use the full likelihood provided by \citep{Ross:2014qpa}.
For the DR11 Ly$\alpha$ data, we use the errors and covariances reported in Ref. \citep{Aubourg:2014yra}.
The correlations of the six BOSS DR12 data points in Ref.~\citep{Alam:2016hwk} are also considered.
The $6\times 6$ covariance matrix for BOSS DR12 is \citep{Alam:2016hwk}
\begin{equation}
    \textbf{Cov}_{\text{DR12}} =
    \begin{pmatrix}
        624.7 & 23.73 & 325.3 & 8.350 & 157.4 & 3.578 \\
        23.73 & 5.609 & 11.64 & 2.340 & 6.393 & 0.968 \\
        325.3 & 11.64 & 905.8 & 29.34 & 515.3 & 14.10 \\
        8.350 & 2.340 & 29.34 & 5.423 & 16.14 & 2.853 \\
        157.4 & 6.393 & 515.3 & 16.14 & 1375.1 & 40.43 \\
        3.578 & 0.968 & 14.10 & 2.853 & 40.43 & 6.259 \\
    \end{pmatrix}.
\end{equation}

\begin{table*}
    \centering
    \caption{BAO measurements used in this work. The sound horizon size of the fiducial model is $r_{s,\text{fid}}=147.78$ Mpc in Ref.~\citep{Alam:2016hwk}.}
    \label{tab:BAO}
    \begin{tabular}{lccccccc}
        \hline
        \hline
        Data-set & $z_{\text{eff}}$ & Observable & Measurement & Reference \\
        \hline
        6dFGS & $0.106$ & $r_s/D_V$ & $0.327 \pm 0.015$ & \citep{Beutler:2011hx} \\
        \hline
        SDSS DR7 MGS & $0.15$ & $D_V/r_s$ & $4.466 \pm 0.168$ & \citep{Ross:2014qpa} \\
        \hline
        \multirow{6}*{BOSS DR12} & 0.38 & $D_M(r_{s,\text{fid}}/r_s)$ [Mpc] & $1512.39 \pm 24.99$ &  \multirow{6}*{\citep{Alam:2016hwk}} \\
        ~ & 0.38 & $H(r_s/r_{s,\text{fid}})$ [km/s/Mpc] & $81.21 \pm 2.37$ & ~ \\
        ~ & 0.51 & $D_M(r_{s,\text{fid}}/r_s)$ [Mpc] & $1975.22 \pm 30.10$ \\
        ~ & 0.51 & $H(r_s/r_{s,\text{fid}})$ [km/s/Mpc] & $90.9029 \pm 2.33$ \\
        ~ & 0.61 & $D_M(r_{s,\text{fid}}/r_s)$ [Mpc] & $2306.68 \pm 37.08$ \\
        ~ & 0.61 & $H(r_s/r_{s,\text{fid}})$ [km/s/Mpc] & $98.9647 \pm 2.50$ \\        
        \hline
        eBOSS DR14 & $1.52$ & $D_V/r_s$ & $26.47 \pm 1.23$ & \citep{Hou:2018yny} \\
        \hline
        BOSS DR12 Ly$\alpha$ forest & $2.33$ & $(c/H)^{0.7} D_M^{0.3}/r_s$ & $13.94\pm 0.35$ & \citep{Bautista:2017zgn} \\
        \hline
        \multirow{2}*{BOSS DR11 Ly$\alpha$ QSO} & $2.36$ & $c/(H r_s)$ & $9.0\pm 0.3$ & \multirow{2}*{\citep{Font-Ribera:2013wce}} \\
        ~& $2.36$ & $D_A/r_s$ & $10.8 \pm 0.4$ & ~ \\
        \hline
        \hline
    \end{tabular}
\end{table*}

In Table.~\ref{tab:BAO}, the observable $D_V \equiv \left[ c z D_M(z)^2 /H(z) \right]^{1/3}$ is the volume average distance.
Here $c$ is the light speed.
The parameter $r_s$ is the comoving sound horizon at the end of radiation drag epoch $z_d$, shortly after recombination, when baryons decouple from the photons:
\begin{equation}
    r_s = \int_{z_d}^{\infty} \frac{c_s(z)}{H(z)} dz,
    \label{eq:rd}
\end{equation}
where $c_s(z)$ is the sound speed of the photon-baryon fluid.
In this work, we only consider the kinematic property of the late evolution of the universe but not the cosmic component in the cosmography approach.
There are no definitions of parameters for the early physics, such as the energy densities of baryons and radiation, and the sound speed $c_s(z)$.
Thus, the parameter $r_s$ is treated as a free parameter in our paper  \footnote{For the \lcdm model, we only consider the late evolution of the universe without involving early evolution, so $r_s$ is also considered as a free parameter.}.

The sound horizon $r_s$ can be accurately determined by CMB experiments such as Planck and WMAP.
The latest constrained value of $r_s$ can be found in the 2018 Planck Legacy Archive (PLA) tables \footnote{\url{https://wiki.cosmos.esa.int/planck-legacy-archive/index.php/Cosmological_Parameters}}
\begin{align}
    & r_s = 147.05\pm 0.30 {\rm Mpc}, & {\rm Planck~2018}, \label{eq:rdP18}\\
    & r_s = 148.5 \pm 1.2 {\rm Mpc}, & {\rm WMAP~9}, \label{eq:rdW9}
\end{align}
where the Planck value (P18) is obtained from the likelihood combination TT+TE+EE+lowE, and the nine-year WMAP estimate value (W9) is also provided in PLA tables.

The data is combined into a $\chi^2$ statistic
\begin{align}
	\chi^2_{\text{BAO}} &= \chi^2_{\text{6dF}} + \chi^2_{\text{MGS}} + \chi^2_{\text{DR12}} +\chi^2_{\text{DR14}} \nonumber \\
	&~ +\chi^2_{\text{DR12Ly}\alpha} + \chi^2_{\text{DR11Ly}\alpha},
\end{align}
and the six $\chi^2_i$ are all given in the form of $\chi^2_i = (\textbf{w}_i-\textbf{d}_i)^T \cdot \textbf{Cov}_i^{-1} \cdot (\textbf{w}_i -\textbf{d}_i)$.
  Here, the vector $\textbf{d}_i$ is the observational data of the $i$-th type data-set from Table.~\ref{tab:BAO}, $\textbf{w}_i$ is the prediction for these vectors in a given cosmological model, and $\textbf{Cov}_i$ is the covariance matrix of different BAO data-set.

\subsection{Sampling method and priors of free parameters}
\label{sec:prior}

The global constraints on the cosmographic parameters, .i.e., $\{H_0, q_0, j_0, s_0, l_0\}$, the spatial curvature $\omk$ and the sound horizon $r_s$ are performed using the Markov Chain Monte Carlo (MCMC) sampling method.
It's easy to do this by using the publicly available code \textbf{Cobaya}~\footnote{\url{https://github.com/CobayaSampler/cobaya}}, which calls the MCMC sampler developed for CosmoMC \citep{Lewis:2002ah,Lewis:2013hha}.
In order to put constraint on the free parameters, we have calculated the overall likelihood $\mathcal{L} \propto \exp[-\chi^2/2]$, where $\chi^2$ can be defined by
\begin{equation}
    \chi^2 = \chi^2_\text{CC} + \chi^2_{\text{Pantheon}} +\chi^2_\text{BAO}.
\end{equation}

Priors are needed in order to explore the posteriors of the free parameters.
We impose uniform prior on the free parameters, with prior ranges listed in table~\ref{tab:priors}.
In our calculations, to ensure the physical meaning of observable, we should artificially guarantee the positiveness of $H(z)$, $D_M(z)$, $D_L(z)$ and $D_A(z)$, by setting the posterior to be zero once any of them turn out negative.

\begin{table}
    \centering
    \caption{The priors for the model parameters.}
    \label{tab:priors}
    \begin{tabular}{cc}
        \hline
        Parameters & Priors \\
        \hline
        {\boldmath $H_0$} & [50, 90] \\
        {\boldmath $q_0$} & [-2, 0.0] \\
        {\boldmath $j_0$} & [-10, 10] \\
        {\boldmath $s_0$} & [-150, 150] \\
        {\boldmath $l_0$} & [-1000, 1000] \\
        {\boldmath $\omk$} & [-1, 1] \\
        {\boldmath $r_s$} & [130, 160] \\
        \hline
    \end{tabular}
\end{table}

\section{constraint results and analysis}
\label{sec:results}

In this section, we describe the observational constraint on the cosmography models using various cosmic observational data-set summarized in section~\ref{sec:obs_meth}.
In particular, we focus on the Hubble constant $H_0$ and the spatial curvature parameter $\omk$ in order to investigate whether the spatial curvature can relax the $H_0$ tension problem. 

\subsection{Constraint results under different expansion truncation}
\label{sec:result_order}

\begin{figure*}
    \centering 
    \includegraphics[scale=0.5]{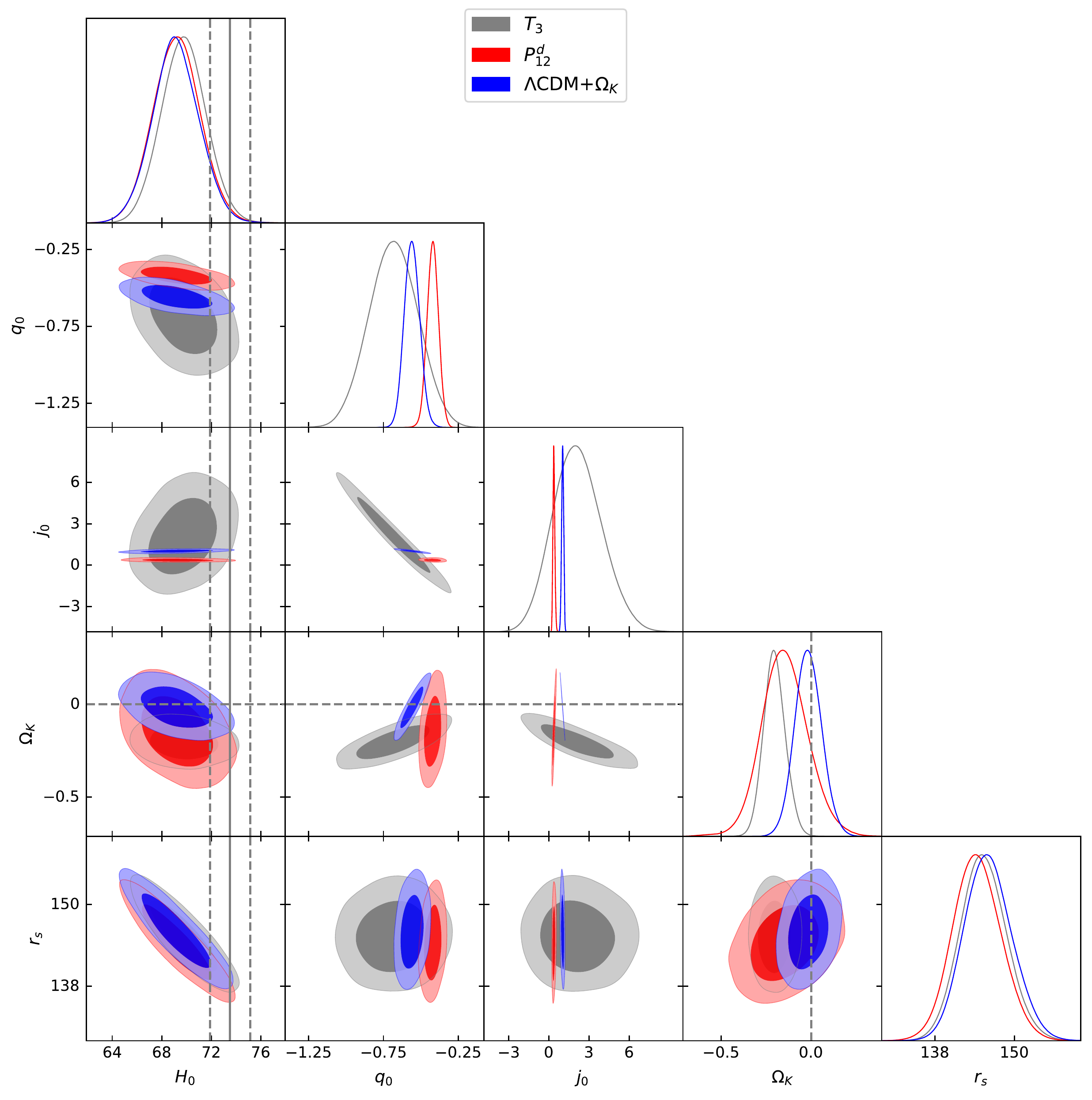}
    \caption{One-dimensional and two-dimensional marginalized distributions with $1\sigma$ and $2\sigma$ contours for the selected parameters under $T_3$, $P_{12}^d$, and \lcdmk models. Horizontal and vertical lines in the $H_0$-related plots indicate the mean value (solid lines) and $1\sigma$ confidence limits (dashed lines) of R18. The dashed lines in $\omk$-related plots represent $\omk = 0$.}
    \label{fig:T3P12Lcdm}
\end{figure*}

\begin{figure*}
    \centering 
    \includegraphics[scale=0.5]{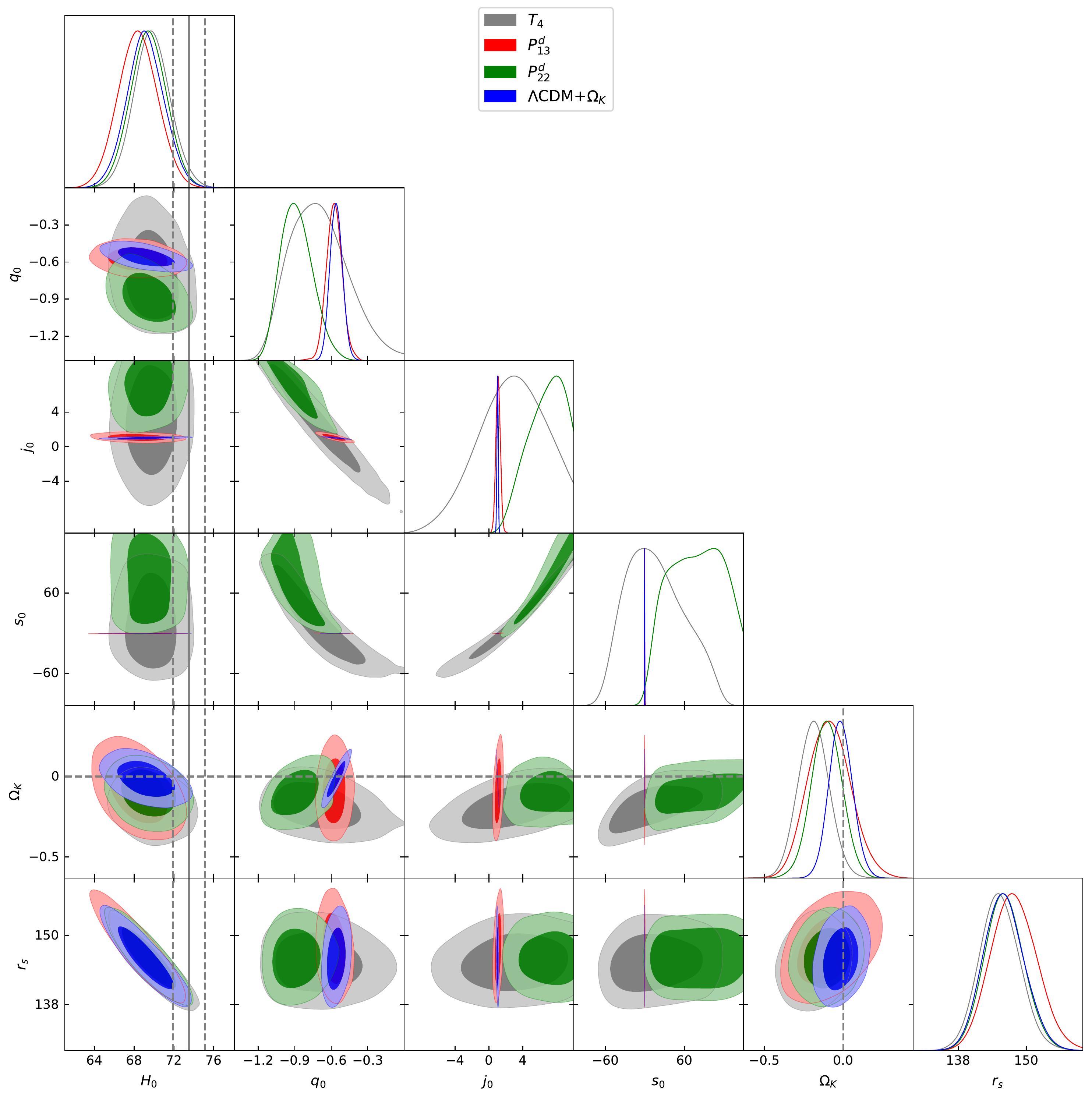}
    \caption{One-dimensional and two-dimensional marginalized distributions with $1\sigma$ and $2\sigma$ contours for the selected parameters under $T_4$, $P_{22}^d$, $P_{13}^d$, and \lcdmk models. Horizontal and vertical lines in the $H_0$-related plots indicate the mean value (solid lines) and $1\sigma$ confidence limits (dashed lines) of R18. The dashed lines in $\omk$-related plots represent $\omk = 0$.}
    \label{fig:T4P22P13}
\end{figure*}

\begin{figure*}
    \centering 
    \includegraphics[scale=0.5]{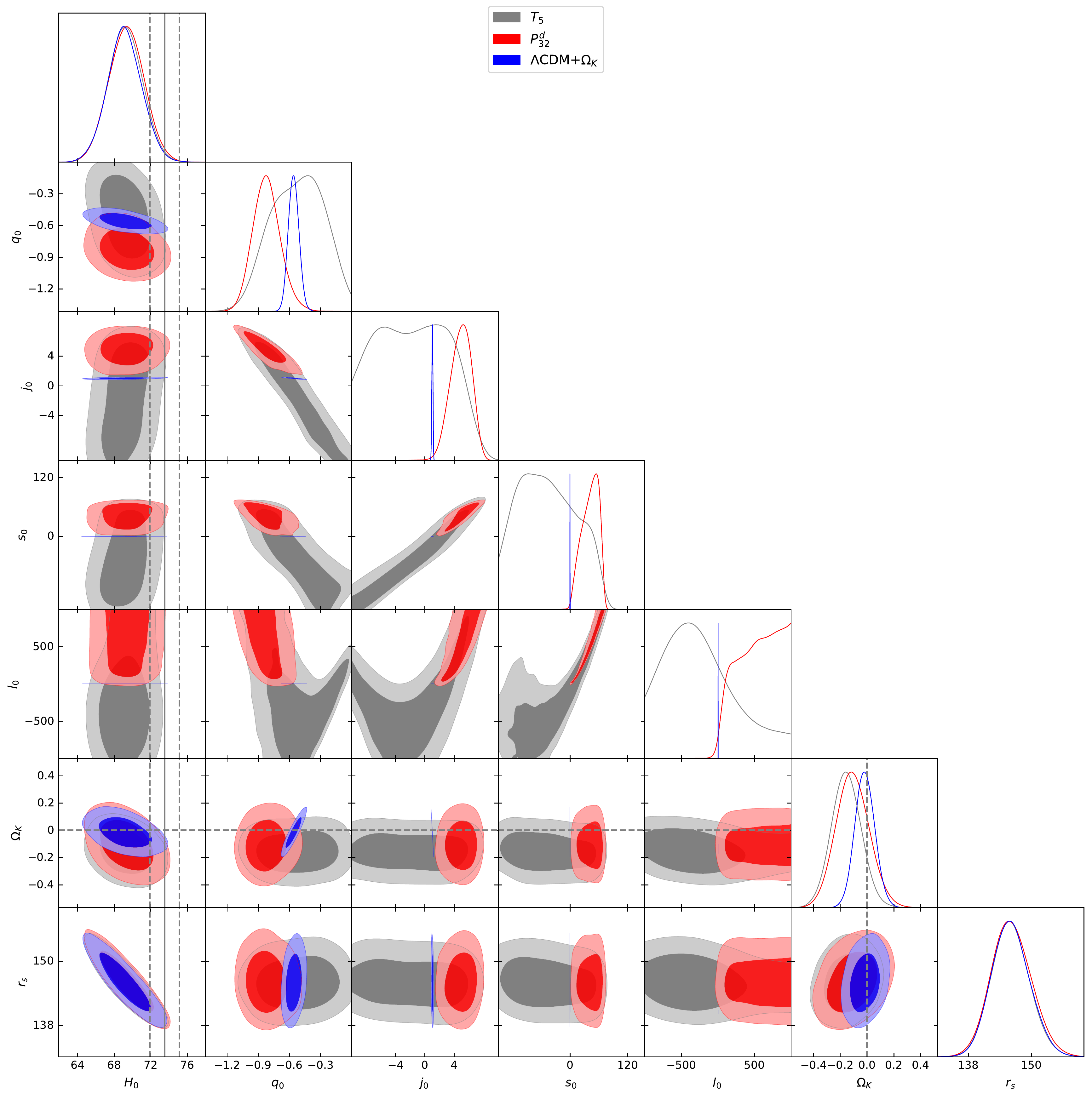}
    \caption{One-dimensional and two-dimensional marginalized distributions with $1\sigma$ and $2\sigma$ contours for the selected parameters under $T_5$, $P_{32}^d$, and \lcdmk models. Horizontal and vertical lines in the $H_0$-related plots indicate the mean value (solid lines) and $1\sigma$ confidence limits (dashed lines) of R18. The dashed lines in $\omk$-related plots represent $\omk = 0$.}
    \label{fig:T5P32}
\end{figure*}

Let us first focus on the performance of the cosmography models with different expansion truncation (or the number of free model parameters) under the constraint of all the mentioned cosmic observational data points.
The mean and $68.3\%$ confidence limits are summarized in Table~\ref{tab:result_sr}. 
Figures~\ref{fig:T3P12Lcdm} - \ref{fig:T5P32} show the likelihood distributions of model parameters of approximants up to the $j_0$, $s_0$, and $l_0$ terms, respectively.
For comparison, the contours of \lcdmk model have been shown in all three figures.

Contour plots in Figs.~\ref{fig:T3P12Lcdm}, \ref{fig:T4P22P13}, and \ref{fig:T5P32} show that the degeneracies among the parameters in the Pad\'{e} approximants are similar as that of the Taylor series in terms of $y$.
The constraint results show that $H_0$ and $r_s$ vary slightly among different expansion orders, but $q_0$, $j_0$, $s_0$, $l_0$ and $\omk$ change a lot between different approximants.

\begin{table*}
    \centering
    \caption{Constrained cosmographic parameters by the CC+Pantheon+BAO data-set under different expansion orders within $1\sigma$ confidence level. Here ``---'' denotes there is no constraint result.}
    \label{tab:result_sr}
    \renewcommand\arraystretch{1.2}
    \begin{tabular}{ccccccccccc}
        \hline
        \hline
        Model & $H_0$ & $q_0$ & $j_0$ & $s_0$ & $l_0$& $\omk$ & $r_s$ \\
        \hline
        $P_{12}^d$ & 
        $69.2\pm 1.9$ & $-0.420\pm 0.038$ & $0.365^{+0.059}_{-0.078}$ & --- & --- & $-0.15^{+0.12}_{-0.13}$ & $144.4\pm 3.6$ \\
        $P_{13}^d$ &
        $68.4\pm 2.0$ & $-0.575 \pm 0.065$ & $1.08\pm 0.26$ & $-0.29^{+0.16}_{-0.31}$ & --- & $-0.09\pm 0.13$ & $147.8\pm 4.0$ \\
        $P_{22}^d$ &
        $69.4\pm 1.8$ & $-0.88^{+0.11}_{-0.15}$ & $6.6^{+3.0}_{-1.2}$ & $78^{+50}_{-40}$ & --- & $-0.105\pm 0.096$ & $146.1\pm 3.5$ \\
        $P_{32}^d$ &
        $69.3\pm 1.9$ & $-0.82^{+0.12}_{-0.14}$ & $4.9^{+1.6}_{-1.3}$ & $42^{+20}_{-10}$ & $l_0 > 392$ & $-0.11\pm 0.12$ & $146.2^{+3.5}_{-4.0}$ \\
        \hline
        $T_3$ &
        $69.8\pm 1.8$ & $-0.68\pm 0.16$ & $2.1^{+1.7}_{-1.9}$ & --- & --- & $-0.204^{+0.056}_{-0.063}$ & $145.4\pm 3.5$ \\
        $T_4$ &
        $69.8\pm 1.8$ & $-0.70^{+0.20}_{-0.28}$ & $2.6^{+4.5}_{-3.6}$ & $17^{+34}_{-56}$ & --- & $-0.187\pm 0.098$ & $145.3\pm 3.5$ \\
        $T_5$ &
        $69.2\pm 1.9$ & $-0.53^{+0.28}_{-0.25}$ & $-1.6\pm 4.6$ & $-45^{+47}_{-79}$ & $-228^{+280}_{-630}$ & $ -0.15\pm 0.11$ & $146.0\pm 3.6$ \\
        \hline
        \lcdmk & 
        $69.1\pm 1.9$ & $-0.559\pm 0.051$ & $1.016\pm 0.074$ & $-0.325^{+0.081}_{-0.10}$ & $3.16\pm 0.21$ & $-0.016\pm 0.074$ & $146.1^{+3.3}_{-3.8}$ \\
        \hline
        \hline
    \end{tabular}
\end{table*}

From the constraint results, one can find that $r_s$ measured using different approximants are consistent with the values of P18 and W9.
From the $r_s -H_0$ contours, it is easy to find that $r_s$ is anti-correlated with $H_0$. 
This anti-correlation can be explained.
Considering Eq.~(\ref{eq:rd}), it can be found that if we fix the evolutionary history of the universe, then a larger $H_0$ would result in a smaller $r_s$.
Moreover, the increase of $H_0$ could reduce the angular diameter distance $D_A$, and a reduction in $r_s$ can lead to the same BAO measurements of $r_s/D_V$ or $Hr_s$.
Therefore, any new physics proposed in order to explain the tension between the CMB estimation of $H_0$ and the direct measurements needs to reduce the sound horizon in the early universe.
In addition, $r_s$ has no physical definition and is just set as a free parameter, thus, $r_s$ can only be constrained by the BAO data-points in this paper.
As a result, we find that the variance of $r_s$ in different models are much bigger than that of P18 and W9.

The values listed in Table~\ref{tab:result_sr} show that the best-fit of $H_0$ are consistent with the \lcdmk results and a little lower than the R18 value by $1.5-2.0\sigma$ in both the Taylor series and the Pad\'e approximants.
This is similar with the results presented in Refs.~\citep{Gomez-Valent:2018hwc, Gomez-Valent:2018gvm}, where they have used the similar data combination but adopted different model-independent reconstruction techniques.
Moreover, our results of $H_0$ are also consistent with that in the non-flat \lcdm, XCDM and $\phi$CDM models under almost the same data combinations \citep{Park:2018tgj}.

It is noteworthy that the mean value of $\omk$ is negative in all the considered cosmography models and a closed universe is favored at more than $1\sigma$ significance in the $P_{12}^s$, $P_{22}^d$, $T_3$, $T_4$, and $T_5$ cases.
This suggests that the data combinations used in this work favor a closed universe.
Moreover, from all the $H_0-\omk$ planes in the three figures we can find that there is a visible anti-correlation between $H_0$ and $\omk$, which indicates that a smaller $\omk$ will result in a larger $H_0$.

The results in Table~\ref{tab:result_sr} show that the constraints on parameters $q_0$ and $j_0$ are consistent with \lcdmk model in Taylor series and $P_{13}^d$ cases, but have more than $2\sigma$ tension with $P_{12}^d$, $P_{22}^d$, and $P_{32}^d$ cases.
This phenomena is in accord with Fig.~\ref{fig:pade_dh}, where the Taylor series and $P_{13}^d$ cases can give good approximations to the \lcdm model under the same model parameters.
The contour plots in Figs.~\ref{fig:T3P12Lcdm} - \ref{fig:T5P32} also show that there are strong correlations between different cosmographic parameters, which suggest that other data-set is needed to break these parameters' degeneracies.
Moreover, the present observational data can not tightly constrain $s_0$ or $l_0$ in all the adopted approximants.

\subsection{Bayesian evidence }
\label{sec:bayesian}

Which model will be the most favored one by the data-set used in the present paper or how many series terms do we need to include in the model functions?
We need a statistical comparison among all the series expansion models.
Bayesian evidence is a good measure of the statistical preference for a model over another one by computing the Bayes factor \citep{Trotta:2008qt}.
In this section, we apply the Bayesian evidence method to determine which model is most favored by the observational data.

For a given model $\mathcal{M}$ with a parameter space $\theta$, and a specific observational data $d$, the Bayesian evidence $E$ is defined as
\begin{equation}
    E = p(d| \mathcal{M}) = \int p(d|\theta, \mathcal{M}) \pi(\theta|\mathcal{M}) d\theta,
    \label{eq:bayesE}
\end{equation}
where $\pi(\theta | \mathcal{M})$ is the prior of $\theta$ in model $\mathcal{M}$, and $p(d| \theta, \mathcal{M})$ is the likelihood.
Then, for the two models $\mathcal{M}_i$ and $\mathcal{M}_j$, combining the Bayes' theorem the posterior probability is
\begin{align}
    \frac{p(\mathcal{M}_i|d)}{p(\mathcal{M}_j|d)} = \frac{p(d|\mathcal{M}_i)}{p(d|\mathcal{M}_i)} \frac{\pi(\mathcal{M}_i)}{\pi(\mathcal{M}_i)} = B_{ij} \frac{\pi(\mathcal{M}_i)}{\pi(\mathcal{M}_i)}.
    \label{eq:bayesF}
\end{align}
where $B_{ij}$ is the Bayes factor of the model $\mathcal{M}_i$ relative to $\mathcal{M}_j$.
The strength of the preference for one of the competing models over the other is usually determined by means of the Jeffreys scale \citep{Kass:1995loi, Trotta:2008qt} (see Table~\ref{tab:jeffreys}).
 
\begin{table}
    \centering
    \caption{Revised Jeffreys scale quantifying the observational viability of any model $\mathcal{M}_i$ compared to some reference model $\mathcal{M}_j$.}
    \label{tab:jeffreys}
    \begin{tabular}{ccc}
        \hline
        \hline
        $\ln B_{ij}$ & Strength of evidence for model $\mathcal{M}_i$ \\
        \hline
        $0< |\ln{B_{ij}}|<1$ & Weak \\
        $1< |\ln{B_{ij}}|<3$ & Definite/Positive \\
        $3< |\ln{B_{ij}}|<5$ & Strong \\
        $|\ln{B_{ij}}| > 5$ & Very Strong \\
        \hline
        \hline
    \end{tabular}
\end{table}

Here we apply the publicly available code \textbf{MCEvidence}\footnote{\url{https://github.com/yabebalFantaye/MCEvidence}} to calculate the logarithm of the Bayes factor for different models.
\textbf{MCEvidence} can directly calculate the Bayesian evolving from MCMC chains \citep{Heavens:2017hkr, Heavens:2017afc}.

In Table~\ref{tab:Bij}, we have shown the values of $\ln B_{ij}$ of all the cosmography models in this paper.
The negative values of $\ln B_{ij}$ indicate that \lcdmk is the most preferred model among all the models we have considered.
From the table, we see a weak evidence for $T_3$ model against \lcdmk, and positive evidence for $T_4$, $T_5$, and $P_{22}^d$ cases.
And the other Pad\'{e} approximants are strongly disfavored by the cosmic observational data compared with the \lcdmk model.

\begin{table}
    \centering
    \caption{The values of $\ln B_{ij}$ computed for the selected cosmography model ($\mathcal{M}_i$), where the reference scenario is the \lcdmk model ($\mathcal{M}_j$).}
    \label{tab:Bij}
    \begin{tabular}{ccc}
        \hline
        \hline
        Model & $\ln B_{ij}$ &Evidence against \lcdmk \\ 
        \hline
        $P_{12}^d$ & $-7.85$ & Very Strong \\ 
        $P_{13}^d$ & $-9.19$ & Very Strong \\ 
        $P_{22}^d$ & $-2.19$ & Definite/Positive \\ 
        $P_{32}^d$ & $-6.67$ & Very Strong \\ 
        \hline
        $T_3$ & $-0.14$ & Weak \\ 
        $T_4$ & $-1.97$ & Definite/Positive \\ 
        $T_5$ & $-2.27$ & Definite/Positive \\ 
        \hline
        \hline
    \end{tabular}
\end{table}

\subsection{Different priors of $H_0$ and $r_s$}
\label{sec:sub_prior}

The constraint results in Table~\ref{tab:result_sr} show that the estimations of $H_0$ are closer to the \textit{Planck} \lcdm extimation of $H_0$ than that of SH0ES.
However, we should note that the previous analysis did not consider the effects of the latest value from SH0ES collaboration R18 and the precise determinations of $r_s$ from PLA, i.e., P18 or W9.
As we know, there is a $3.8 \sigma$ difference between the local distance ladder and \textit{Planck} estimations of $H_0$ \citep{Riess:2018byc}, and the values of sound horizon $r_s$ from \textit{Planck} and WMAP are also different.
Thus, we will use the importance sampling technique \citep{Lewis:2002ah} to investigate the influences of the three different priors, i.e., gaussian priors with dispersions as given in R18, P18, and W9, upon the whole parameter space.

\begin{table}
    \centering
    \caption{Constrained cosmographic parameters using CC+Pantheon+BAO data combination within $1\sigma$ confidence level by adopting different priors in the \lcdmk model.}
    \label{tab:lcdm_prior}
    \begin{tabular}{cccccccc}
        \hline
        \hline
        Parameters & R18 & P18 & W9\\ 
        \hline
        $H_0$ & $72.10^{+0.82}_{-1.1}$ & $68.66\pm 0.85$ & $68.12\pm 0.91$\\ 
        $q_0$ & $-0.603\pm 0.041$ & $-0.554\pm 0.047$ & $-0.552^{+0.046}_{-0.052}$\\ 
        $j_0$ & $1.068\pm 0.062$ & $1.008\pm 0.069$ & $1.005\pm 0.071$\\
        $\omm$ & $0.310\pm 0.022$ & $0.303\pm 0.023$ & $0.302^{+0.023}_{-0.021}$ \\
        $\omk$ & $-0.068\pm 0.062$ & $-0.008\pm 0.069$ & $-0.005\pm 0.071$\\ 
        $r_s$ & $141.3\pm 2.1$ & $147.05\pm 0.25$ & $148.28^{+0.92}_{-1.1}$\\         
        \hline
        \hline
    \end{tabular}
\end{table}

\begin{table}
    \centering
    \caption{Constrained cosmographic parameters using CC+Pantheon+BAO data combination within $1\sigma$ confidence level by adopting different priors in the $P_{22}^d$ approximant.}
    \label{tab:pade_prior}
    \begin{tabular}{cccccccc}
        \hline
        \hline
        Parameters & R18 & P18 & W9\\ 
        \hline
        $H_0$ & $71.96^{+0.97}_{-1.1}$ & $68.97\pm 0.89$ & $68.5\pm 1.0$\\ 
        $q_0$ & $-0.94^{+0.11}_{-0.14}$ & $-0.87^{+0.12}_{-0.15}$ & $-0.87^{+0.12}_{-0.15}$\\ 
        $j_0$ & $j_0 > 5.92$ & $6.5^{+3.4}_{-1.0}$ & $6.5^{+3.2}_{-1.2}$\\ 
        $s_0$ & $76\pm 40$ & $76^{+50}_{-40}$ & $77^{+50}_{-40}$ \\
        $\omk$ & $-0.137\pm 0.092$ & $-0.106\pm 0.096$ & $-0.102\pm 0.097$\\ 
        $r_s$ & $142.0\pm 2.3$ & $147.04\pm 0.27$ & $148.3\pm 1.0$\\         
        \hline
        \hline
    \end{tabular}
\end{table}

\begin{table}
    \centering
    \caption{Constrained cosmographic parameters using CC+Pantheon+BAO data combination within $1\sigma$ confidence level by adopting different priors in the $T_3$ approach.}
    \label{tab:taylor_prior}
    \begin{tabular}{cccccccc}
        \hline
        \hline
        Parameters & R18 & P18 & W9\\ 
        \hline
        $H_0$ & $72.20^{+0.88}_{-1.1}$ & $68.99^{+0.85}_{-0.94}$ & $68.48\pm 0.98$\\ 
        $q_0$ & $-0.76^{+0.14}_{-0.16}$ & $-0.67\pm 0.17$ & $-0.67\pm 0.16$\\ 
        $j_0$ & $2.8\pm 1.8$ & $2.0^{+1.7}_{-1.9}$ & $2.0^{+1.7}_{-2.0}$\\
        $\omk$ & $-0.213^{+0.053}_{-0.059}$ & $-0.20^{+0.055}_{-0.064}$ & $-0.202\pm 0.061$\\ 
        $r_s$ & $141.5\pm 2.3$ & $147.05\pm 0.25$ & $148.24^{+0.94}_{-1.1}$\\         
        \hline
        \hline
    \end{tabular}
\end{table}

The importance sampling results of the three models, i.e., \lcdmk, $P_{22}^d$ and $T_3$, are listed in Tables~\ref{tab:lcdm_prior}, \ref{tab:pade_prior}, and \ref{tab:taylor_prior}.
Comparing the data in the above three tables with Table~\ref{tab:result_sr}, one can find that different priors on $r_s$ have little effect on the cosmographic parameters, but the R18 $H_0$ depresses $q_0$ and $\omk$ while magnifies $j_0$ due to the degeneracies between them.
Note that the best-fit values of $r_s$ listed in Table~\ref{tab:result_sr} are similar to the prior of P18 or W9 while the best-fit values of $H_0$ are very different from R18, so it is easy to understand why the P18 and W9 priors have little influence on the cosmographic parameters but R18 prior has great influences on them.
Specially, we also find that a closed universe is supported at more than $1\sigma$ confidence level by the three models with R18 prior on $H_0$.
Therefore, we know that more precise direct or indirect determinations of the magnitude of the spatial curvature in the future will be helpful in resolving the $H_0$ tension problem.


\section{Conclusions}
\label{sec:conclusions}

In this paper, adopting the Hubble parameter data, SNe Ia data, and the BAO data, we have investigated the spatial curvature parameter and the cosmographic parameters via the model-independent cosmography approach.
To overcome the convergence problem, we have used two methods: one is Taylor series of $D_M(z)$ in terms of $y=z/(1+z)$, the other is using the Pad\'{e} approximant method.
In order to figure out which method could give the better approximation, we compare them with the \lcdmk model.
Finally, we find that Taylor series in terms of $y$ up to 3-rd, 4-th and 5-th orders all can give better approximation of \lcdm model, and when using the Pad\'{e} approximant method, $P_{12}^d$, $P_{13}^d$, $P_{22}^d$ and $P_{32}$ give better approximations of the \lcdmk model.
Then, in this paper, using the CC+Pantheon+BAO data-set, we have investigated the three Taylor series models and the four Pad\'{e} approximant models from the 3-rd order to the 5-th order.

We find that the sound horizon $r_s$ is anti-correlated with $H_0$, which suggests that new physics in the early universe capable of lowering the sound horizon will be helpful in relaxing the $H_0$ tension problem.
Besides, $H_0 -\omk$ planes show that the Hubble constant is anti-correlated with the spatial curvature, which points out another way to weaken the $H_0$ tension.
From our constraint results, we find that all the approximants and the \lcdmk model prefer a lower $H_0$ than R18 value. 
The constraint results listed in Table~\ref{tab:result_sr} suggest that a closed universe is preferred by all the Pad\'e and Taylor series approximations.
Finally, adopting the Bayesian evidence method, we find that there is weak evidence for $T_3$ approximants against \lcdmk model and in this case a closed universe is favored at $3.6\sigma$ significance.

One should note that, because there is no definition of the photon-baryon fluid's sound speed in cosmography model, the sound horizon $r_s$ has been treated as a free parameter in the present paper.
Thus, taking the P18 or W9 values as priors on $r_s$ is natural.
These two priors are used for importance sampling on the MCMC chains that have already been obtained, and we find that different priors on $r_s$ have little effects on all the cosmographic parameters and $\omk$.
When we adopt the value of R18 as prior of $H_0$ to do importance sampling, all the best-fit of the cosmographic parameters will change due to the parameter degeneracies.
Still, the results show that a closed universe is supported at more than $1\sigma$ confidence level by the current observational data.

\section*{Acknowledgments}
    
This work is supported in part by National Natural Science Foundation of China under Grant No. 11675032 (People's Republic of China).







\appendix



\section{Formulas used for approximating dimensionless distance}
\label{sec:app_A}

In this Appendix we give the formulas for the approximants of the dimensionless transverse comoving distance used to fit the data, for every Pad\'{e} approximant considered in this work.


The Pad\'{e} approximant of $H_0 D_M(z)/c$ up to 5-th degree:
\begin{align}
    P_{12}^d &= \frac{z}{1 +\frac{1}{2} g_1 z -\frac{1}{12} g_4 z^2}, \label{eq:P12d}\\
    P_{21}^d &= \frac{z +\frac{1}{6} g_4/g_1 z^2}{1 + \frac{1}{3} g_2 z/g_1 z }, \label{eq:P21d}\\
    P_{13}^d &= \frac{z}{1 + \frac{1}{2} g_1 z -\frac{1}{12} g_4 z^2 +\frac{1}{24} g_5 z^3 }, \label{eq:P13d} \\
    P_{22}^d &= \frac{ z + \frac{1}{2} g_5/ g_4 z^2 }{1 + \frac{1}{2} g_7/g_4z + \frac{1}{12} g_8 /g_4 z^2}, \label{eq:P22d} \\
    P_{31}^d &= \frac{z + \frac{1}{4} g_7/ g_2 z^2 -\frac{1}{24} g_8 /g_2 z^3}{1 + \frac{1}{4} g_3 /g_2 z}, \label{eq:P31d} \\
    P_{14}^d &= \frac{z}{1 + \frac{1}{2} g_1 z -\frac{1}{12} g_4 z^2 +\frac{1}{24} g_5 z^3 -\frac{1}{720} g_6 z^4}, \label{eq:P14d} \\
    P_{23}^d &= \frac{ z +\frac{1}{30} g_6/ g_5 z^2 }{1 + \frac{1}{60} f_4/ g_5 z^2-\frac{1}{360} f_5/ g_5 z^3+ \frac{1}{30} f_6/ g_5 z}, \label{eq:P23d} \\
    P_{32}^d &= \frac{z + \frac{1}{10} f_4/ g_8 z^2 +\frac{1}{60} f_5/ g_8 z^3}{1 + \frac{1}{5} f_2/g_8 z +\frac{1}{20} f_3/g_8 z^2}, \label{eq:P32d} \\
    P_{41}^d &= \frac{z + \frac{1}{10} f_1/ g_3 z^2 -\frac{1}{30} f_2/ g_3 z^3 +\frac{1}{120} f_3/g_3 z^4 }{1 + \frac{1}{5} g_9/ g_3 z}, \label{eq:P41d}
\end{align}
where
\begin{align}
    g_1 &= 1+q_0, \\
    g_2 & = 2 - j_0 + 4 q_0 + 3 q_0^2 + \omk, \\
    g_3 &= 6 + 18 q_0 + 27 q_0^2 + 15 q_0^3 - j_0 (9 + 10 q_0) - s_0 + 
    6 (1 + q_0) \omk, \\
    g_4 &= 1 - 2 j_0 + 2 q_0 + 3 q_0^2 + 2 \omk, \\
    g_5 &= 1 + 3 q_0 + 8 q_0^2 + 6 q_0^3 - j_0 (5 + 6 q_0) - s_0 + 
    2 (1 + q_0) \omk, \\
    g_6 &= 19 + 40 j_0^2 - 6 l_0 + 76 q_0 + 286 q_0^2 + 420 q_0^3 + 225 q_0^4 \nonumber \\
    &~ - 2 j_0 (86 + 205 q_0 + 150 q_0^2) - 66 s_0 - 60 q_0 s_0 - 20 (-2 + j_0 \nonumber \\
    &~ - 4 q_0 - 3 q_0^2) \omk - 14 \omk^2, \\
    g_7 &= 2 + 6 q_0 + 13 q_0^2 + 9 q_0^3 - j_0 (7 + 8 q_0) - s_0 + 
    4 (1 + q_0) \omk, \\
    g_8 &= 2 - 4 j_0^2 + 8 q_0 + 23 q_0^2 + 30 q_0^3 + 9 q_0^4 - j_0 (11 + 25 q_0 + 6 q_0^2)\nonumber \\
    &~ - 3 s_0 - 3 q_0 s_0 + (2 + 8 j_0 + 4 q_0 - 6 q_0^2) \omk - 4 \omk^2 \\
    g_9 &= 24 + 10 j_0^2 - l_0 + 96 q_0 + 216 q_0^2 + 240 q_0^3 + 105 q_0^4 -j_0 (72 \nonumber \\
    &~ + 160 q_0 + 105 q_0^2) - 16 s_0 - 15 q_0 s_0 + 5 (7 - 2 j_0 + 14 q_0 \nonumber \\
    &~ + 9 q_0^2) \omk + \omk^2 \\
    f_1 &= 18 + 20 j_0^2 - 2 l_0 + 72 q_0 + 207 q_0^2 + 270 q_0^3 + 135 q_0^4 \nonumber \\
    &~ - j_0 (99 + 225 q_0 + 160 q_0^2) - 27 s_0 - 25 q_0 s_0 - 20 (-2 + j_0 \nonumber \\
    &~ - 4 q_0 - 3 q_0^2) \omk + 2 \omk^2 \\
    f_2 &= 12 - 3 l_0 + 60 q_0 - 3 l_0 q_0 + 216 q_0^2 + 408 q_0^3 + 330 q_0^4 + 90 q_0^5 \nonumber \\
    &~ - 5 j_0^2 (3 + 4 q_0) - 38 s_0 - 73 q_0 s_0 - 30 q_0^2 s_0 - j_0 (96 + 326 q_0 \nonumber \\
    &~ + 325 q_0^2 + 90 q_0^3 + 5 s_0) + 5 (3 + 9 q_0 - 6 q_0^3 + j_0 (9 + 10 q_0) \nonumber \\
    &~ + s_0) \omk - 27 (1 + q_0) \omk^2 \\
    f_3 &= 12 - 40 j_0^3 - 8 l_0 + 72 q_0 - 16 l_0 q_0 + 312 q_0^2 - 12 l_0 q_0^2 \nonumber \\
    &~ + 768 q_0^3 + 927 q_0^4 + 510 q_0^5 + 135 q_0^6 + j_0^2 (-37 - 100 q_0 + 40 q_0^2) \nonumber \\
    &~ - 68 s_0 - 196 q_0 s_0 - 162 q_0^2 s_0 - 30 q_0^3 s_0 - 5 s_0^2 + 2 j_0 (-66 + 2 l_0 \nonumber \\
    &~ - 298 q_0 - 449 q_0^2 - 255 q_0^3 - 90 q_0^4 - 13 s_0 - 20 q_0 s_0) + 4 (4 \nonumber \\
    &~ + 20 j_0^2 - l_0 + 16 q_0 + 16 q_0^2 + 15 q_0^4 + j_0 (8 + 15 q_0 - 30 q_0^2) \nonumber \\
    &~ - s_0) \omk - 4 (8 + 11 j_0 + 16 q_0 - 3 q_0^2) \omk^2 + 4 \omk^3 \\
    f_4 &= 14 - 6 l_0 + 70 q_0 - 6 l_0 q_0 + 277 q_0^2 + 551 q_0^3 + 465 q_0^4 + 135 q_0^5 \nonumber \\
    &~ - 10 j_0^2 (1 + 2 q_0) - 61 s_0 - 116 q_0 s_0 - 45 q_0^2 s_0 - j_0 (137 \nonumber \\
    &~ + 472 q_0 + 495 q_0^2 + 150 q_0^3 + 10 s_0) + 10 (2 + 6 q_0 + q_0^2 - 3 q_0^3 \nonumber \\
    &~ + j_0 (5 + 6 q_0) + s_0) \omk - 34 (1 + q_0) \omk^2 \\
    f_5 &= 4 - 80 j_0^3 - 6 l_0 + 24 q_0 - 12 l_0 q_0 + 120 q_0^2 - 18 l_0 q_0^2 + 320 q_0^3 \nonumber \\
    &~ + 423 q_0^4 + 270 q_0^5 + 135 q_0^6 + 9 j_0^2 (1 + 20 q_0^2) - 36 s_0 - 102 q_0 s_0 \nonumber \\
    &~ - 78 q_0^2 s_0 - 15 s_0^2 + 6 j_0 (-10 + 2 l_0 - 78 q_0^2 - 55 q_0^3 - 45 q_0^4 \nonumber \\
    &~ - 3 s_0 - 2 q_0 (23 + 5 s_0)) + 6 (3 + 20 j_0^2 - 2 l_0 + 12 q_0 + 42 q_0^2 \nonumber \\
    &~ + 60 q_0^3 + 45 q_0^4 - 2 j_0 (12 + 30 q_0 + 35 q_0^2) - 12 s_0 - 10 q_0 s_0) \omk \nonumber \\
    &~ + 6 (1 - 2 j_0 + 2 q_0 + 3 q_0^2) \omk^2 - 28 \omk^3, \\
    f_6 &= 34 + 40 j_0^2 - 6 l_0 + 136 q_0 + 451 q_0^2 + 630 q_0^3 + 315 q_0^4 \nonumber \\
    &~ - j_0 (247 + 575 q_0 + 390 q_0^2) - 81 s_0 - 75 q_0 s_0 \nonumber \\
    &~ + 10 (7 - 2 j_0 + 14 q_0 + 9 q_0^2) \omk - 14 \omk^2.
\end{align}

The dimensionless distance expanded in terms of $y = \frac{z}{1+z}$ up to fifth degree is given by
\begin{align}
    d(y) &= y + \frac{1}{2} (1-q_0) y^2 + \frac{1}{6} (2-j_0-2 q_0+3 q_0^2+ \omk) y^3 \nonumber \\
    &~ +\frac{1}{24} \bigg[ 6-j_0 (3-10 q_0)-6 q_0+9 q_0^2-15 q_0^3+s_0 \nonumber \\
    &~ +6 (1-q_0) \omk \bigg] y^4 + \frac{1}{120} \bigg[ 24+10 j_0^2-l_0-24 q_0+36 q_0^2 \nonumber \\
    &~ -60 q_0^3+105 q_0^4-j_0 (12-40 q_0+105 q_0^2)+4 s_0 \nonumber \\
    &~ -15 q_0 s_0-5 (-7+2 j_0+10 q_0-9 q_0^2) \omk+\omk^2 \bigg] y^5.
    \label{eq:T5y}
\end{align}

\bsp	
\label{lastpage}
\end{document}